\documentclass[12pt]{article}
\usepackage[letterpaper]{geometry}

\usepackage{amssymb,amsmath,amsfonts,eurosym,float,geometry,ulem,graphicx,caption,color,setspace,sectsty,comment,footmisc,caption,natbib,pdflscape,subfigure,array,sectsty, titlesec}

\sectionfont{\fontsize{14}{12}\selectfont}
\titleformat*{\section}{\large\bfseries}
\titleformat*{\subsection}{\normalsize\bfseries}

\newcolumntype{L}[1]{>{\raggedright\let\newline\\arraybackslash\hspace{0pt}}m{#1}}
\newcolumntype{C}[1]{>{\centering\let\newline\\arraybackslash\hspace{0pt}}m{#1}}
\newcolumntype{R}[1]{>{\raggedleft\let\newline\\arraybackslash\hspace{0pt}}m{#1}}

\usepackage{natbib}
\bibpunct[, ]{(}{)}{,}{a}{}{,}%
\def\bibfont{\small}%
\def\bibsep{\smallskipamount}%
\usepackage[hidelinks]{hyperref}
\usepackage{titlesec}
\titlespacing\section{0pt}{0pt plus 2pt minus 4pt}{0pt plus 2pt minus 2pt}
\usepackage{caption}
\usepackage{indentfirst}

\begin{document}
\raggedright
\setlength{\parindent}{1cm}
	\begin{titlepage}
		\title{Should we trust web scraped data?}
		\author{ Jens Foerderer  \\ Technical University of Munich  \\ \small \href{mailto:jens.foerderer@tum.de}{\nolinkurl{jens.foerderer@tum.de}} \\
        }
        \date{\small Working paper, comments welcome \\ \today }
		\maketitle
		\begin{abstract}
            \noindent\justify
            The increasing adoption of econometric and machine-learning approaches by empirical researchers has led to a widespread use of one data collection method: web scraping. Web scraping refers to the use of automated computer programs to access websites and download their content. The key argument of this paper is that naïve web scraping procedures can lead to sampling bias in the collected data. This article describes three sources of sampling bias in web-scraped data. More specifically, sampling bias emerges from web content being volatile (i.e., being subject to change), personalized (i.e., presented in response to request characteristics), and unindexed (i.e., abundance of a population register). In a series of examples, I illustrate the prevalence and magnitude of sampling bias. To support researchers and reviewers, this paper provides recommendations on anticipating, detecting, and overcoming sampling bias in web-scraped data.
            %\noindent Placeholder\\
            %\vspace{0in}\\
            %\noindent\textbf{Keywords:} key1, key2, key3\\
            %\vspace{0in}\\
            %\noindent\textbf{JEL Codes:} key1, key2, key3\\
            \bigskip
\end{abstract}
\setcounter{page}{0}
\thispagestyle{empty}
\end{titlepage}
	
	%\pagebreak 
	\doublespacing

\section{Introduction}
The rapid increase in the use of econometric, machine learning, natural language processing, and artificial intelligence methods empirical research has given particular prominence to one data collection method: web scraping \citep[e.g.,][]{Agarwal.2014, Boegershausen.2022, Einav.2014, Landers.2016, Salganik.2017}. Web scraping refers to the use of automated computer programs to download content from websites, such as forum posts, product details, or user profiles. Web scraping opens the avenue to accessing novel data, and at an unprecedented scale \citep{Greene.2022, Lin.2013}. The past years have seen an extraordinary increase in the use of this data collection method across social science disciplines, as I will review below.

The core argument of this paper is a cautionary one: web-scraped data can show considerable sampling bias. The sources of the sampling bias originate in the properties of web content, namely being volatile (i.e., being subject to change), personalized (i.e., presented in response to request characteristics), and unindexed (i.e., abundance of a population register). Naïve web scraping procedures that do not account for these properties can show considerable sampling bias. In this article, I describe each characteristic and illustrate the prevalence and magnitude of sampling bias in real-world web-scraped data. To support researchers and reviewers, I provide recommendations regarding when to expect sampling bias, how to test for it, and strategies to overcome.\\
By structuring these insights, this paper addresses an important gap. As of now – researchers and reviewers are left to rely on their own idiosyncratic reasoning to assess the validity of web-scraped data. This contrasts with the sizable progress made in the past decades to understanding biases in data collected via surveys or interviews \citep[e.g.,][]{Groves.2009, Podsakoff.2003}. While widespread, web scraping has so far been evaluated mostly along technical, ethical, and legal considerations \citep{Boegershausen.2022, Landers.2016}. Some few empirical studies that rely on web scraped data elaborate on sampling bias in particular data collections \citep[e.g.,][]{alsudais2021incorrect,Cavallo.2017,sen2021total}, yet a holistic treatment of this issue is absent. \\
The paper proceeds as follows. Chapter 2 provides a background on web scraping and sampling bias. Chapter 3 proceeds by outlining sources of sampling bias, illustrating each with real-world data, and providing recommendations for anticipating, detecting, and addressing them. Chapter 4 concludes.
\section{Background: Data Collection with Web-Scraping}
Web scraping describes the automated process of accessing websites and downloading their content \citep{Boegershausen.2022, Edelman.2012}. Web content is a rich source of data for empirical research \citep[e.g.,][]{Golder.2014, Grover.2020, Lazer.2017}. Such content includes, for example, posts, likes, comments, and followers on social media platforms such as Facebook, or Twitter. Another example of such data are product prices, seller strategies, and customer reviews in online marketplaces. Web scrapers can be implemented by programming a script from scratch, or relying on pre-built tools, including Requests, Selenium, and ScrapeHub \citep{Mitchell.April2018}.\footnote{Application programming interfaces (APIs) provide a structured access to a web data source. APIs typically permit sending queries to a backend database, such that researchers can define precisely what data they need and such that the data is returned in a structured format. Although the queries are also sent via automated software programs, data collection via APIs requires a different process than web scraping. Therefore, data collection via APIs is not covered in this article.} 

\begin{figure}[] % H erzwingt den Ort der Abbildung
        \caption{\raggedright\label{fig:scraping} Web Scraping Process: From the Target Population to the Sample} 
        \footnotesize Web scraping entails two steps, \textit{indexing} and \textit{fetching}. In indexing, the target population is systematically registered. Indexing yields the frame in terms of a register of all units in the population, together with the URLs pointing to each unit. Fetching automatically visits each URL listed in the frame and downloads the resource at which it points, typically an HTML document.
        \newline\newline
		\includegraphics[width=1.0\textwidth]{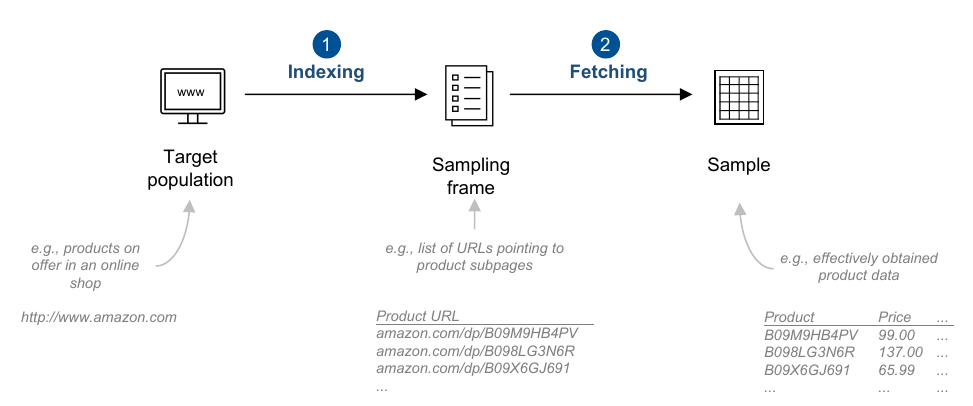}
\end{figure}

Web scraping entails two steps, \textit{indexing} and \textit{fetching}, as displayed in Figure \ref{fig:scraping}. First, the target population is systematically indexed. The population is comprised of units. For example, a researcher might be interested in studying the target population of products listed in an online shop, whereby the products represent the units of the population. Indexing yields the \textit{sampling frame} in terms of a register of all units in the population. The sampling frame covers all units in the target population and is used to construct the sample. In practice, the researcher compiles a set of URLs pointing to websites that shall be downloaded. In the traditional survey approach, this step is comparable to a researcher obtaining a population register from which, subsequently, the units can be drawn for the sample \citep{Groves.2009}. For web scraping, there are different approaches to obtain an index. Perhaps the most straightforward approach is that the website displays an index of all units. For example, a researcher seeking to collect data on all products offered in an online shop might discover that the shop provides a product catalogue.\\
Second, the URLs contained in the sampling frame are fetched. The scraper visits each URL and downloads the resource at which it points, typically an HTML document. The resource is then parsed in terms of not needed clutter stripped (e.g., HTML or JavaScript code), the meaningful intended data is compiled into a dataset, and erroneous entries removed, which represents the effective sample that enters the study.

Web scraping has made it possible to considerably advance our understanding of several core questions in the information systems field, including online sales \citep[e.g.,][]{Datta.2018, ghose2014estimating}, participation on online platforms \citep[e.g.,][]{Burtch.2013, Huang.2016}, revenue models \citep[e.g.,][]{yan2022does, liu2021digital}, cyber attacks and fraud \citep[e.g.,][]{crosignani2023pirates, Luca.2016}, and online-offline interactions \citep[e.g.,][]{Forman.2009, Greenwood.2016}. Beyond, it has been applied to understand economic phenomena, including the formation of prices \cite{Cavallo.2017}, corruption \citep{Callen.2015}, economic uncertainty \citep{Altig.2020}, media bias and slant \citep{Garz.2020}, and job mobility \citep{BoydSwan.2018}. In the political sciences, it has been used to investigate censorship \citep{King.2013}, partisanship \citep{OSMUNDSEN.2021}, and collective memory \citep{FOUKA.2023}. Further, web data provided novel answers to research questions in psychology, including personality judgments \citep{Youyou.2015} and sentiment formation \citep{Zhang.2021}. It has also enabled unique insights into sociology, including regarding social movements \citep{Vasi.2015}, policing \citep{Cheng.2022}, and gender biases \citep{Leung.2018}.\footnote{Scraping has been applied when the collaboration with a website owner is not possible or would violate research goals, or when the platform does not have an interface for data collection (i.e., an API). This raises various ethical and legal questions that are discussed elsewhere (see Boegershausen et al. 2022).}

Of crucial interest for empirical research is to what degree the collected sample data represents the target population \citep{Bhattacherjee.2012, Groves.2009}. The \textit{sampling error} captures how well a sample estimate generalizes to the target population. A sampling error exists when there are differences between the characteristics of the sample and of the target population. An error is not necessarily an issue for inference as long as the source of it is random. If the probability of sample inclusion is known for each unit of the population, valid inference is possible via statistical adjustment. Therefore, sampling strategies with the goal of representation implement – at least to some degree – random sampling. By randomly selecting units from the population, each population unit has an identical chance of being selected. The error then has an asymptotic mean of zero, so the random error does not affect the mean of a variable. Sampling error interferes with the estimation when it is systematic (i.e., non-random). Any systematic error that occurs because some units of the population are more likely to be sampled than others creates invalid inference and is therefore also referred to as a \textit{sampling bias}.
\section{Sources of Sampling Bias in Web-scraped Data}
In the following, this paper presents three sources of sampling bias in web-scraped data. These sources relate to the characteristics of web content, namely being volatile, personalized, and unindexed.
\subsection{Volatility}
\textbf{Description}

\begin{figure}[] % H erzwingt den Ort der Abbildung
	%\begin{flushleft}
        \caption{\raggedright\label{fig:volatility} How Volatility causes Sampling Bias} 
        \footnotesize (A) illustrates how volatility causes sampling bias. The horizontal axis shows the passage of calendar time, and the vertical axis separates the population units. The length of the arrow indicates the time that a population unit was observable on the website. The researcher scrapes at time $S$ with the goal of obtaining all population units. Population units not captured in the sample are marked red. (B) reports what percentage of the true data was obtained assuming the scraping would have taken place in t + 10 Mins, t + 1 Hour, t + 8 Hours, t + 1 Day, t + 7 Days, or t + 14 Days. Red marks the unobserved share. In (C), the markers report the results of a test for the difference in means between the true data and each of the scraped samples. Rating is the number of likes minus the dislikes of an answer as voted by other users. Answer Length is the total number of characters of the answer. The marker indicates the estimate and whiskers give 95\% confidence intervals.
        \newline\newline
		%{\footnotesize}
		\includegraphics[width=1.0\textwidth]{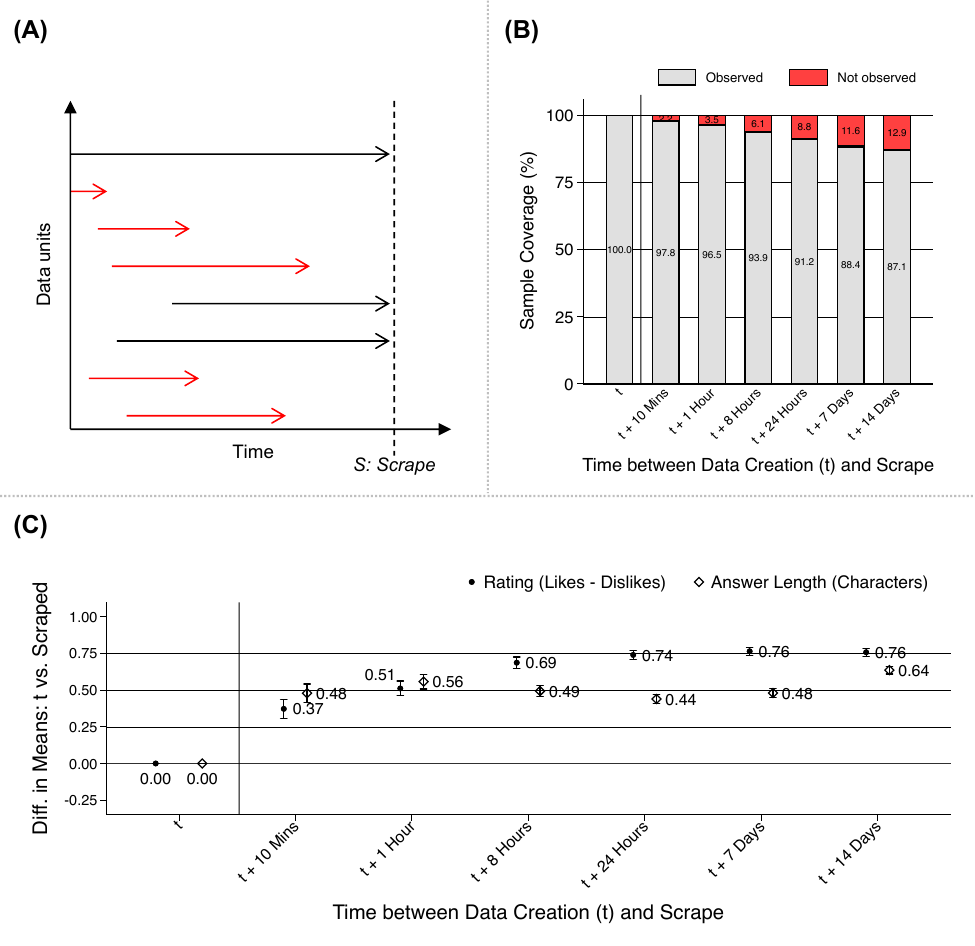}
	%\end{flushleft}
\end{figure}

Websites and their content are volatile \citep[e.g.,][]{Golder.2014,ruths2014social}. They change over time. Website owners keep their content current and relevant. The data can be adapted to cover new information, reflect changes in the real world, or provide up-to-date data. Websites of businesses, organizations, or institutions are updated to reflect changes in their operations, offerings, staff profiles, prices, or contact information. Platforms that allow user-generated content, such as forums, comment sections, or social media, are inherently volatile. Users can contribute, modify, or delete content. User-generated content is also subject to moderation and censorship \citep{King.2013,liu2022implications}. As a consequence, population units are seldom permanently available but, instead, they appear and disappear.\\
Figure \ref{fig:volatility} (A) illustrates how volatility causes sampling bias in web-scraped data. When data is volatile, the time gap between when the data was created and when it is scraped becomes crucial: a web scraper can only capture the data which is still available at the time of the scrape but not at the time of data creation. As a consequence, the probability of being sampled correlates with the length of time that the web content is available. The horizontal axis shows the passage of calendar time, and the vertical axis separates different population units that are to be scraped. The population units, denoted by the arrows, are available for a certain period of time as indicated by the length of the arrow. The researcher seeks to collect all of the population units, and scrapes at time $S$. Units that disappeared before $S$ are not captured in the dataset, marked as red. Volatility creates a sampling bias such that the sample contains only the data still available at time $S$, marked black. The basic logic is similar to a length or a survival bias \citep[e.g.,][]{vanEs.2000}.\\
\textbf{Example}\\
As an example, consider data that we obtained from an online question-and-answer platform similar to StackOverflow or Quora. The platform almost exclusively relies on user-generated content. Users post questions, and other users reply (“answers”), whereby the most helpful answer is decided by voting. Data from such settings has been extensively used in prior research, for instance, to understand users' motivation to contribute or what makes users contribute better content. The platform is moderated; designated moderators delete superficial answers or those that violate platform guidelines. The data comprises all answers that were posted over the two-week period between June 1 and June 14, 2023, directly from the platform operator. The data contains the exact timestamp when a post was created and when it was removed. The data used here is the \textit{true} data as stored in the database of the platform operator. I then calculate what share of answers could be captured by a web scraper depending on when the scraping takes place relative to the time the answer was posted $t$ (i.e., depending on the time lag between when an answer was posted and when the scraper collected it).\\
Figure \ref{fig:volatility} (B) shows that there are considerable discrepancies between the scraped and the true data, even when the time lag between data creation and scraping is short. The bars indicate what percentage of the true data was obtained by the web scraper assuming the scraping would have taken place in t + 10 Mins, t + 1 Hour, t + 8 Hours, t + 1 Day, t + 7 Days, or t + 14 Days, whereby t denotes the time the answer was published. In this very setting, volatility is marked. If scraping takes place only 10 minutes after the answer was posted, the scraper fails to collect 2.1\% of the answers. If the lag amounts to eight hours, the scraper fails to collect 5.4\%.\\
Figure \ref{fig:volatility} (C) shows that the web-scraped data shows sampling bias. The horizontal axis displays the different samples, with the true data being marked as t, and the samples scraped with a time lag next. The vertical axis reports the results of a test for the difference in means between the true data and each of the scraped samples. The test is conducted for two exemplary variables that we can observe. \textit{Rating} is the number of likes minus the dislikes of an answer as voted by platform users. \textit{Answer Length} is the total number of characters of the answer. It is evident that the answers in the scraped data are rated much better than in the true data. Perhaps the moderators eliminated low-rated answers swiftly, or their authors withdrew them quickly. The answers in the scraped samples are also significantly longer, perhaps due to similar reasons. Regardless of the precise reasons causing the volatility, the web scraped data shows a sampling bias: it comprises answers that are much better and longer than they actually were in the true data. The researchers encounters biased data.\\

\textbf{Recommendations}\\
The volatility of web data depends on the empirical setting. Online markets - such as for products, jobs, classifieds, or real estate - are an example where content is volatile due to the fluctuating availability of offers. Researchers should be aware that the availability of products will likely correlate with their quality. If a researcher intends to scrape all offers posted over a specific time frame, a web scraper will likely fail to identify the absolute top offers if not adjusted for volatility. For example, in the case of apartment ads, those for high-quality apartments will be taken offline sooner because they quickly attracted many requests, thus motivating a realtor to stop the ad. Other settings can be less volatile, or not at all. Consider for instance a website that reports historical economic indicators. This type of web data is less volatile, the entry of data is relatively predictable, and we would not expect data exit to take place. \\
Researchers should particularly evaluate whether volatility correlates with the characteristics of the population units they intend to study. Consider for example social media platforms such as Twitter, YouTube, or Instagram. On these platforms content is user-generated and moderated. The moderation filters spam, malicious content, or low-quality user-generated content. Some content is also removed for violating platform policies, for instance, due to hateful speech or being off-topic. Also, censorship might be present. Overall, this degree of volatility can have a sizeable consequence on what data can be scraped. For example, \cite{Timoneda.2018} studied politically charged Tweets and found that between 2 to 2.5\% of them are removed within 15 minutes after publication.\\
Researchers should also be aware that volatility also arises from purely technical reasons. Some websites frequently change their structure, such as the URL path structure or the HTML source code structure. Web scrapers are amenable to structural changes in the target website and might then break \citep[e.g.,][]{Datta.2018}. This has two consequences. First, researchers might not be aware that the scraper failure is linked to volatility, and therefore underestimate that it might cause sampling bias. Second, it requires researchers to adapt the scraper to handle the new structure. Such fixes take time, and the result is that a time lag between data creation and scraping accrues.\\
To detect volatility, researchers are advised to carry out a pilot test before the actual data collection. The idea of this test is to understand whether population units exit over time and, if so, how quickly they disappear. Researchers should create a test sampling frame that contains the URLs of some population units, obtained immediately after these population units were created. These URLs should then be queried in regular time intervals to assess if the URLs are still available. For example, in the case of the question-and-answer platform, a researcher could create a test sampling frame of answers, obtained immediately after they were published. Then, every minute, the researcher tests if the answer can still be retrieved via its URL. The collected data effectively permits replicating Figure \ref{fig:volatility} (B) and (C), and therefore understanding to what degree volatility causes sampling bias. \\
To account for volatility, researchers should reduce the time lag between when a population unit is created and when it is scraped. For example, the web scraper can be equipped with a monitoring component, which controls changes made to a website \citep{mallawaarachchi2020change}. Once the monitor detects a change, the scraping should be triggered. Alternatively, if monitoring is not possible or puts too much burden on the target, web scrapers should scrape data in highly frequent time intervals.\\
If the degree of volatility is known, researchers can implement reweighting approaches. The before-outlined pilot test permits determining which units are more likely to disappear than others based on their characteristics. \cite{marconi2022content} describes a two-step procedure, namely first fitting a survival model to estimate the hazard rate at which web data disappears, and then using the obtained parameters to weight the population units. To counterbalance volatility, units with a higher estimated hazard rate are given a relatively higher weight.

\subsection{Personalization}
\textbf{Description}

\begin{figure}[] % H erzwingt den Ort der Abbildung
	%\begin{center}
		\caption{\raggedright\label{fig:personalization} How Personalization Causes Sampling Bias} \footnotesize (A) shows the setup. A set of identical web scrapers requests search results from an online video platform. The only exception is that scrapers I - V rely on different HTTP request headers, whereby one field of the header was altered respectively. In (B), the bars show how many of the units of the benchmark configurations could be observed when scraping with the particular profile. Non-coverage is marked red. In (C), we compare the configurations with the benchmark on two exemplary variables. Views are the number of views that a video receives and log-adjusted for skewness. Duration captures the length of a video in seconds, also log-adjusted for skewness. The marker indicates the estimate and whiskers give 95\% confidence intervals.
        \newline\newline
		%{\footnotesize }
		\includegraphics[width=1.0\textwidth]{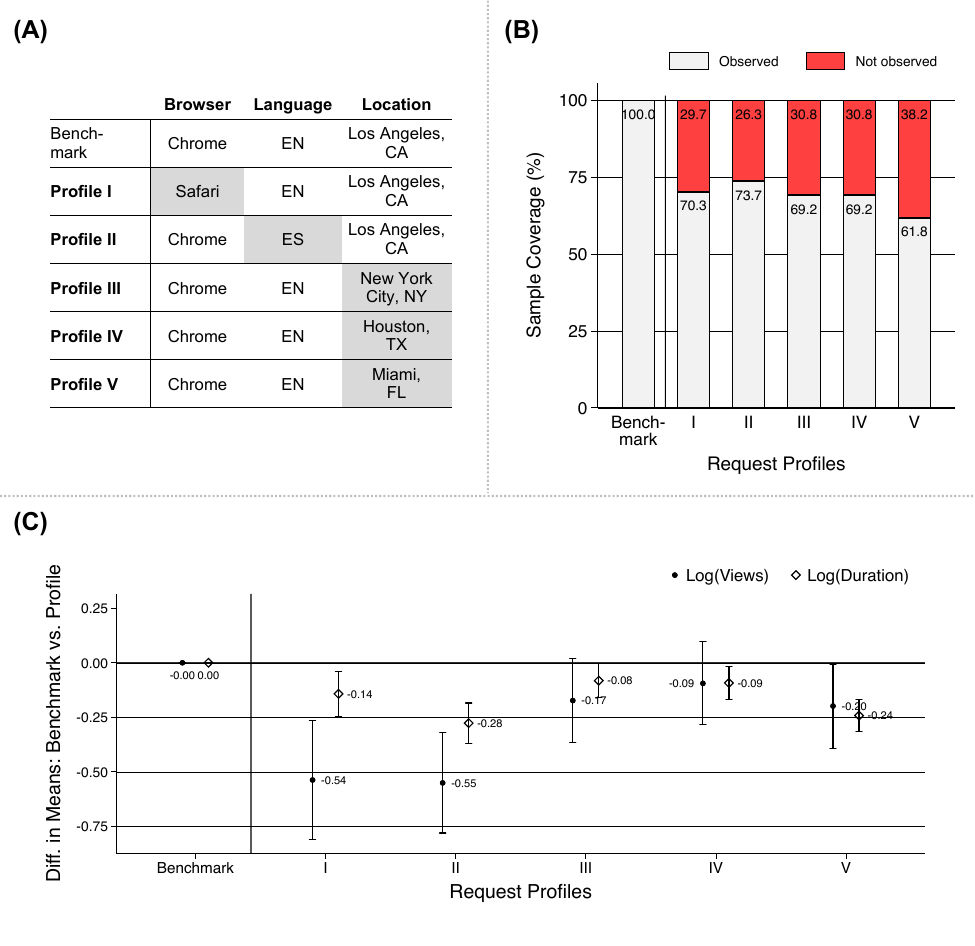}
	%\end{center}
\end{figure}

Much web content is personalized \citep[e.g.,][]{awad2006personalization,Ghose.2014,hosanagar2014will,lambrecht2013does,yoganarasimhan2020search}. A website might display different data to one visitor than to another. Personalization aims to tailor content to individual visitors based on their preferences. Websites personalize content display by incorporating meta-information about the visitor. Such meta-information is available from the so-called HTTP-header that is attached to every request to the web server. For example, the HTTP-header contains information about the browser (e.g., Chrome, Safari) and operating system (e.g., Windows, MacOS) used, the language settings specified in the browser (e.g., English, Spanish), or the IP address, which permits inferring the coarse location from where the request originates. Personalization also incorporates information stored in tracking cookies \citep[e.g.,][]{Cavallo.2017, Laub.3202023}. Tracking cookies might reveal browsing and purchasing histories. Websites also 
personalize content to users who have registered accounts. When users create an account on a website, they often provide personal information such as their preferences, browsing history, location, and other relevant data. Websites use this information to tailor the content and user experience specifically for each individual. \\
Personalization can introduce a sampling bias into the data. Because the website dynamically adjusts the display of content based on meta-information about the visitor, population units differ in their probability of being displayed to the visitor (i.e., web scraper) and, therefore, in their probability of being included in the sample. Assigned with a higher probability are population units that the website considers – in abstract terms – more $relevant$ to the user. Since the web scraper tries to mimic a real visitor as much as possible, a website will display personalized content as it would to a real visitor. As a result, the displayed content differs depending on the metadata available about the web scraper. As a consequence, researchers encounter data that is skewed toward specific population segments, potentially impacting the validity and representativeness of their analysis. \\
Personalization is important to be considered because some websites exploit even tiny pieces of meta-information to personalize content, and perhaps unknown to the researcher. As I document in the following example, even the choice of browser that is used by the web scraper can cause sampling bias. \\
\textbf{Example}\\
To illustrate, consider data that we collected from an online video sharing platform similar to YouTube or DTube. On the platform, users can operate video channels via which they can provide and monetize video content. Visitors can watch videos, follow channels, and discover new videos through keyword search.\\
Figure \ref{fig:personalization} (A) shows the test setup. The basic idea is to assess if the video platform displays different videos to a web scraper depending on the meta-information available in the HTTP-header that is sent with every request of the scraper. For this purpose, different $profiles$ were set up, whereby each profile altered one field of the HTTP request header \citep{fielding1999hypertext}. The benchmark profile is a web scraper based on Chrome (user-agent), English (accept-language), and a Los Angeles based IP-address (X-forwarded-for). The benchmark profile serves as a comparison – we use it to judge the coverage of population units obtained from the other profiles. In profile I, we alter the browser, namely setting it to Safari. In profile II, we alter the system language and set it to Spanish. In configuration III to V, we alter the location and send the request from New York City, Houston, and Miami (i.e., through the use of proxy IP addresses). The scraper's task was then to enter various keywords into the platform's search interface, and then fetch the returned results.\footnote{For simplicity, the scraper searched for the most frequented search terms on video platforms, namely “music,” “comedy,” “tutorial,” “review,” “sport,” and “gaming” \citep{Statista.2023}, and then aggregated the results across the search terms.}
To avoid temporal differences, we execute each scraper in parallel, i.e. at exactly the same time of the day. \\
Figure \ref{fig:personalization} (B) shows that population coverage differs substantially across profiles. The horizontal axis shows the scraper profiles. The bars tell what share of the units obtained via the benchmark configuration are also observed when scraping with the respective profile. Non-coverage is marked red. Simply by altering the browser, the overlap is only 70.3\% (I). When setting the system language to Spanish, the overlap is at 73.7\% (II). We also observe strong coverage discrepancies based on the location, ranging between 61.8\% and 69.2\% (III-V). In summary, even small changes to the request header, coverage differs substantially.\\
Figure \ref{fig:personalization} (C) shows that the web-scraped data shows statistically significant differences across various characteristics. The figure tests for differences in means between the benchmark and each profile along two exemplary variables. \textit{Views} counts how often a video has been watched and is log-adjusted for skewness. \textit{Duration} captures the length of a video in seconds, also log-adjusted for skewness. There are significant differences between the benchmark and the profiles along these variables. For example, videos displayed to a Safari web scraper have been watched less and are shorter (configuration I). Also, videos displayed to a web scraper in Miami are significantly shorter and have fewer views.\\
\textbf{Recommendations}\\

Before collecting the actual data, researchers should analyze the website for cues of personalization. Researchers should be cautious when data is presented as a result of a search functionality, when there is any indication that the results follow a non-obvious sorting. Typical settings are social media platforms, online marketplaces, and search engines. Some websites and platforms indicate that data is displayed by relevance, which hints at personalization based on the visitor’s profile. Moreover, if a website requires creating a user account or requires tracking cookies, researchers should interpret this as a hint at personalization. In addition, the researcher can inspect the target website from from different browsers, computers, or using a virtual private network (VPN) tool, and assess if the displayed content differs. In this process, researchers should also evaluate if the data source not only alters the display of population units, but also their attributes. Examples are websites that engage in yield management, such as airlines or railroad firms, which display different prices to different users to exploit differences in their willingness to pay for a service \citep{chen2022statistical,hinz2011price}. \\
A more formal approach for testing for sampling bias is to replicate the setup in Figure \ref{fig:personalization} (A). Researchers can scrape several samples via different profiles and then compare the obtained data. For example, to test if an online retailer personalized prices based on location, \cite{Cavallo.2017} scraped the product prices from different IP addresses, and checked whether retailers increased prices if a product page was visited often. Describing the scraping profile that was used in a study, for instance the browser, system language, location, and further metadata, permits later studies to collect data using the identical setup.\\
To overcome personalization, some technical means can be implemented. To avoid that content becomes personalized after recurring visits, web scrapers can be programmed to rotate over different attributes of the http header (i.e., to change the user-agent or IP address), or to employ fake cookies \citep{lawson2015web}.

\subsection{Unindexed}
\begin{figure}[] % H erzwingt den Ort der Abbildung
	%\begin{center}
		\caption{\raggedright\label{fig:heuristics} How Incompleteness Causes Sampling Bias}
		%{\footnotesize}
      \footnotesize (A) describes marker and traversing heuristics. In marker-based heuristics, the web scraper scrapes all units associated with the marker m. Some units of the population are not marked, and therefore not sampled (red). In traversing, the subpages are nodes, and hyperlinks connecting the pages are the edges. The scraper travels from s over node to node, and is unable to sample non-connected ones (red). (B) compares the coverage in each of the scraped samples against the true data. Non-coverage is marked red. (C) tests for differences in means between the samples and the true data along two exemplary variables. Satisfaction is the average user rating given for a game on a scale from 0\% to 100\%. Number of reviews is the total number of buyer reviews. The marker indicates the estimate and whiskers give 95\% confidence intervals.
      \newline\newline
		\includegraphics[width=1.0\textwidth]{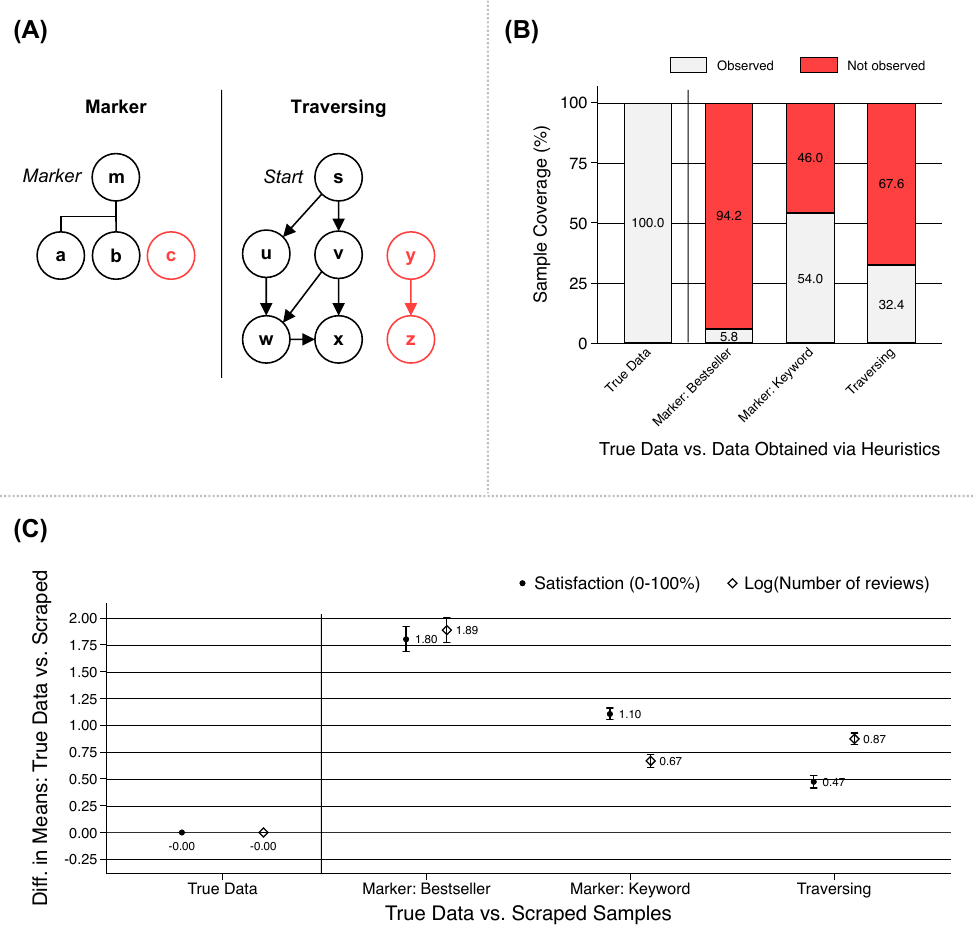}
	%\end{center}
\end{figure}

\textbf{Description}\\
The indexing step of the web scraping procedure is challenged by the fact that obtaining a list of all population units is not trivial, sometimes even impossible \citep[e.g.,][]{edelmann2020computational,kossinets2006effects}. This is because many web data sources do not permit access to the full index and instead display visitors only a subset of the data. For example, commercial websites present only a subsection of their data to prevent competitors from copying the data and launching a rival product. Other websites restrict the display of data for reasons of not overwhelming visitors with \textit{too} much choice. Moreover, compiling a full list might be infeasible because the population is simply too large to be indexed within reasonable time.
\\
In the absence of an index, researchers have resorted to indexing heuristics, two of which are displayed in Figure \ref{fig:heuristics} (A). One heuristic are \textit{markers}, namely to obtain a sampling frame by relying on keywords, hashtags, categories, or other selectors \citep{edelmann2020computational,ruths2014social}. As shown in the figure, the researcher decides for a marker m and obtains the population units (denoted by a node) assigned with the marker m. For example, a researcher interested in studying posts made about politicians in an online forum might resort to construct an index of posts by searching for politicians' names via the forum's search functionality. Another heuristic is \textit{traversing}, meaning that the scraper departs from one part of the website (e.g., the homepage) and walks through the website by following hyperlinks. As shown in the figure, the web scraper begins at a start node s and traverses along the edges (hyperlinks) until all nodes have been discovered \citep{Cothey.2004,kossinets2006effects}. For example, social network samples have been obtained by traversing from one user to that user's followers, and so on \citep[e.g.,][]{gonzalez2014assessing}. It can also be implemented in online marketplaces, where the scraper departs from one product and then follows recommendations for similar products (e.g., “other users purchased these products”).\\
Sampling bias can arise from the use of the heuristics because some nodes are simply not discovered. With marker-based heuristics, only nodes associated with the marker are returned, but this might not include all units of the population (i.e., node \textit{c}). Similarly, traversing-based heuristics might not uncover all units because some units are not connected to the start node (i.e., nodes \textit{y} and \textit{z}). The choice of the start node is subjective; not everything is connected to the start node. Not all subpages of a website can be reached from the start node; this heuristic makes an assumption that the sample can be obtained from the start node. The impact is that units that represent isolated, marginalized portions of the population may not be represented in the sample. Several research papers provide further evidence of sampling biases in datasets constructed via these heuristics \citep[e.g.,][]{cohen2013classifying}.\footnote{A related issue in traversing is that units with more ingoing hyperlinks are more likely to be included, and units with fewer ingoing hyperlinks are less likely to be included. This issue is problematic when scraping underlies constraints, such as if there is reason to believe that procedural errors in the scraping procedure are more likely to occur later in scraping time (e.g., because the scraper is blocked by the website or runs into errors). In such a case, the inclusion of a unit into the sample will correlate with the position (and frequency of occurrence) of a unit based on the website structure.} 
\\
\textbf{Example}\\
To illustrate, consider data obtained from an online marketplace for computer games. The marketplace acts as a storefront from which consumers can discover, review, and purchase games. It lists several hundreds of games. In its basic appearance, it is similar to the Apple App Store, Google Play, or Amazon Marketplace. Assume that a researcher seeks to study the population of racing games. To assess how well different heuristics cover the population, we obtained data in two ways: the \textit{true} data was obtained from the store owner as of June 1, 2023. In parallel, we run several web scrapers that were identical except that each of them implemented a different indexing heuristic. Scraper I relied on a marker heuristic, namely inclusion in the bestseller list. The bestseller list contained the 250 most sold games in the store. Scraper II relied on another marker, namely games returned for the keyword “racing” via the store’s search function. Scraper III implemented a traversing approach by starting with the most purchased racing game and then following stepwise the recommendations for “similar games” on the product detail page.\\
Figure \ref{fig:heuristics} (B) shows that none of the scraping approaches retrieves the entire population. The bars display the coverage of each scraped sample in comparison to the true data. Non-coverage is marked red. Traversing misses 77.4\% of the data; the keyword-based marker approach misses 46.0\% of the population. 

Figure \ref{fig:heuristics} (C) documents the evidence for the sampling bias arising from the use of the heuristics. The scraped samples are compared to the population on two exemplary variables. \textit{Satisfaction} is the average user rating given for a game on a scale from 0\% to 100\%. \textit{Number of reviews} is the total number of reviews submitted by buyers for the game. The scraped set of games in both the marker-based and traversing-based approaches is more well-rated by players and attracts a larger number of reviews. The scraped data in this setting is thus dispersed toward the popular units and lacks coverage of the marginalized units.\\
\textbf{Recommendations}\\
Researchers should except incompleteness whenever a population index is not available. Moreover, although some displayed content might look complete, they, in fact, might not be. Researchers are therefore advised to collaborate with website owners or platforms as much as possible to obtain the index of population units. Unfortunately, collaborating with website owners can be difficult depending on the research question, and raises questions over the replicability of the results \citep{ruths2014social}. In other settings, it is possible to work with professional third parties that have created an index for a website or platform out of commercial interest. These providers combine different sources of data to create an accurate index as much as possible.\\
In some settings, one approach to avoid the use of heuristics can be to reverse-engineer the index by \textit{guessing} in terms of identifying a pattern in unit identifiers and then generating them. Continuing the example of Figure \ref{fig:heuristics}, the researcher might discover that the URLs of the game detail pages follow the same pattern; namely containing a 9-digit identifier for the product. The researcher can then test the possible range of the identifier and reverse-engineer an index. Guessing is, however, not always applicable because the universe of potential identifiers can be complex, and it might still face bias when some units systematically deviate from the identifier format.\\

In other settings, one approach can be cross-validation. The researcher can compare the sample size to publicly available data on the target population. In addition, the researcher can compare aggregated values, as derived from the scraped sample, to overall statistics reported for the population. For example, if other research papers or reliable sources have scraped the data from the target before, a researcher can compare the obtained sample. Also, even if a platform owner is not willing to provide the full index, perhaps the owner is willing to report aggregate numbers that can permit a guess of the coverage. Assuming that website operators cannot share full population data, perhaps they are willing to provide population parameters, including the number of units and the distribution of their characteristics. With such information at hand, researchers can test for statistical differences between the population and the web scraped data. \\
If cross-validation is not feasible, a within-sample validation might be possible. \cite{Duxbury.2021} scraped a darknet drug market, particularly sales and reputation data on sellers. To assess coverage, they compared vendors’ reputations (i.e., the sum of all sales ratings) to manually summing the sales ratings of each transaction in the data.\\
When an index is not available, researchers can follow the approach of \cite{Landers.2016} and develop a \textit{data source theory}. This means that researchers explicitly describe the assumptions that need to be made in order to interpret the scraped data as meaningful. 

\section{Conclusion}
The increasing application of econometric and machine learning methods in empirical research has led to a significant emphasis on web scraping as a data collection approach. Web-scraped data offers access to novel data that can significantly advance research across disciplines. \\
A word of caution emerges in this paper. Web-scraped data can show considerable sampling bias. The sampling bias arises from the nature of web content, in particular its volatile, personalized, and unindexed nature. Researchers employing web scraping need to be aware of these sources. This paper describes each source of bias, and illustrates each of them with real-world examples. If these causes are not addressed, the web scraped data may not accurately reflect the population. \\
This is not to say that web scraped data cannot be trusted; instead, researchers and reviewers are advised to critically reflect on whether and how sampling bias could emerge in a particular setting. The section before has outlined methods for detection as well as technical and statistical recommendations that can be applied to address sampling bias. Moreover, greater transparency over the data collection process, and what measures were taken to overcome sampling bias is needed. This can include information about the websites or pages that were scraped, the selection criteria used, and any exclusions or filtering that was applied. For example, \cite{OSMUNDSEN.2021} describe the results of their web scraping procedure, and errors and provide the programming code. Overall, web scraping data is not a trivial data collection method but requires particular care to yield representative samples.\\

\singlespacing
\setlength\bibsep{0pt}
%\vspace{-0.9cm}
\bibliographystyle{misq}
\bibliography{paper.bib}

\end{document}